\documentclass[11pt,a4paper]{article}
\pdfoutput=1
\usepackage{jheppub}

\usepackage{graphicx,bm}
\usepackage{amsmath,amssymb,mathrsfs,verbatim,subfigure}

\title{Comments on the holographic shear viscosity to entropy density ratio}

\author{Kiminad A. Mamo}

\affiliation{Department of Physics, University of Illinois,
Chicago, IL 60607-7059, USA}

\emailAdd{kabebe2@uic.edu}
\abstract{We revisit the membrane paradigm calculations of the holographic shear viscosity tensor of strongly coupled isotropic plasmas with Einstein gravity dual by emphasizing the fact which was overlooked in the previous literatures that the shear viscosity is a fourth-rank tensor. Using the membrane paradigm we show that depending on whether the holographic shear viscosity tensor to entropy density ratio is $\frac{\eta^{j}\,_{i}\,^{j}\,_{i}}{s}$ or $\frac{\eta^{jiji}}{s}$ or $\frac{\eta_{jiji}}{s}$, we can derive three distinct formulae for the holographic shear viscosity tensor to entropy density ratios given explicitly in terms of the background metric $g_{ij}$. We find that the widely studied $\frac{\eta^{j}\,_{i}\,^{j}\,_{i}}{s}$ holographic shear viscosity tensor to entropy density ratio takes the universal value $\frac{1}{4\pi}$ for isotropic background metric $g_{ij}$ but $\frac{\eta^{jiji}}{s}$ and $\frac{\eta_{jiji}}{s}$ holographic shear viscosity tensor to entropy density ratios take non-universal values which depend on the details of the isotropic background metric $g_{ij}$. %We suggest that even though in most holographic theories the
}


\arxivnumber{}

\begin{document}

\maketitle

\section{\label{sec:introduction}Introduction}
At the center of the application of AdS/CFT techniques \cite{Maldacena:1997re, Gubser:1998bc, Witten:1998qj} for calculating the transport coefficients of strongly coupled plasmas \cite{Son:2002sd, Herzog:2002pc, Policastro:2002se} is the holographic (Kovtun-Son-Starinets) shear viscosity bound which was conjectured by Kovtrun, Son, and Starinets \cite{Kovtun:2004de} after observing the fact that the shear viscosity to entropy density ratio takes the universal value $\frac{1}{4\pi}$ for large class of strongly coupled isotropic quark-gluon plasmas with Einstein (second-derivative) gravity duals \cite{Kovtun:2003wp}. No violation of the bound has been found so far in its regime of validity, i.e., isotropic and Einstein gravity, even though, violation of the bound has been found in anisotropic systems \cite{Rebhan:2011vd, Mamo:2012sy, Polchinski:2012nh, Mateos:2011tv} and higher-derivative gravities \cite{Brigante:2007nu, Myers:2009ij}. In addition, non-universal values have been found in other anisotropic systems \cite{Erdmenger:2010xm, Basu:2009vv}. %In this paper, we report the finding that the bound can actually be violated within its 'regime of validity' in holographic models which %exhaust all the previously studied ones, of course, including the isotropic, strongly coupled $\mathcal{N}=4$ super-Yang-Mills plasma with Einstein gravity %dual.

In this paper we comment on the possibility of finding non-universal holographic shear viscosity to entropy density ratios in strongly coupled isotropic plasmas with Einstein (second-derivative) gravity duals. Our key observation comes from the fact that the boundary energy-momentum tensor operators $T^{\mu\nu}$, $T_{\mu\nu}$, and $T^{\mu}\,_{\nu}$ have different sources in the bulk (i.e., the boundary values of the metric perturbations $h_{\mu\nu}$, $h^{\mu\nu}$, and $h_{\mu}\,^{\nu}$, respectively, which satisfy different equations of motion in the bulk) hence the two-point functions $\langle T^{\mu\nu}T^{\mu\nu}\rangle$,
$\langle T_{\mu\nu}T_{\mu\nu}\rangle$, and
$\langle T^{\mu}\,_{\nu}T^{\mu}\,_{\nu}\rangle$ and the corresponding shear viscosity tensors $\eta^{j}\,_{i}\,^{j}\,_{i}$, $\eta^{jiji}$, and $\eta_{jiji}$, respectively, that we'll extract from them using the Kubo's formulae
\begin{equation}
  \eta^{ijij}=\lim_{\omega\rightarrow 0}\frac{1}{2\omega}\int dtd\mathbf{x}e^{i\omega t}\langle[T^{ij}(x),T^{ij}(0)]\rangle ,
\end{equation}
\begin{equation}
  \eta_{ijij}=\lim_{\omega\rightarrow 0}\frac{1}{2\omega}\int dtd\mathbf{x}e^{i\omega t}\langle[T_{ij}(x),T_{ij}(0)]\rangle ,
\end{equation}
\begin{equation}
 \eta^{i}\,_{j}\,^{i}\,_{j}=\lim_{\omega\rightarrow 0}\frac{1}{2\omega}\int dtd\mathbf{x}e^{i\omega t}\langle[T^{i}\,_{j}(x),T^{i}\,_{j}(0)]\rangle ,
\end{equation}
(where the indices $i$, $j$ stand for the spatial coordinates $x$, $y$, $z$, and $i\neq j$) are different.

This paper is organized as follows: In section \ref{actions}, we write down the effective actions and corresponding equations of motion for the metric perturbations $h_{xy}$, $h^{xy}$, and $h^{x}\,_{y}$.

In section \ref{paradigm}, we use the membrane paradigm to calculate the holographic shear viscosity tensor to entropy density ratios $\frac{\eta^{xyxy}}{s}$, $\frac{\eta_{xyxy}}{s}$, and $\frac{\eta^{x}\,_{y}\,^{x}\,_{y}}{s}$ in terms of the isotropic background metric $g_{ij}$. We find that $\frac{\eta^{xyxy}}{s}$ and $\frac{\eta_{xyxy}}{s}$ are non-universal and depend on the details of the isotropic background metric $g_{ij}$ unlike $\frac{\eta^{x}\,_{y}\,^{x}\,_{y}}{s}$ which always take the universal value $\frac{1}{4\pi}$ independent of the isotropic background metric $g_{ij}$ used.

In section \ref{applications}, we apply the general formulae derived in section \ref{paradigm} for the holographic shear viscosity tensor to entropy density ratios $\frac{\eta^{xyxy}}{s}$, $\frac{\eta_{xyxy}}{s}$, and $\frac{\eta^{x}\,_{y}\,^{x}\,_{y}}{s}$ to the $\mathcal{N}=4$ super-Yang-Mills, ABJM, and (2,0) superconformal plasmas on flat spaces, and show that the non-universal ones $\frac{\eta^{xyxy}}{s}$ and $\frac{\eta_{xyxy}}{s}$ depend explicitly on the temperature $T$ and AdS radius $R$. And, we argue that due to their dependence on the AdS radius R which has no physical interpretation for gauge theories on flat spaces, the non-universal shear viscosity to entropy density ratios $\frac{\eta^{xyxy}}{s}$ and $\frac{\eta_{xyxy}}{s}$ should be discarded as unphysical. However, we point out that, even though, the non-universal shear viscosity to entropy density ratios $\frac{\eta^{xyxy}}{s}$ and $\frac{\eta_{xyxy}}{s}$ have no physical meaning for gauge theories on flat spaces, they might still have some physical implications for gauge theories on compact spaces (like spheres with radius R).

\section{\label{actions}Actions and equations of motion}
The Einstein-Hilbert action
\begin{equation}\label{bulkk}
S_{\rm bulk}=-\frac{1}{2\kappa^2}\int d^{5}x\sqrt{-g}\,R,
\end{equation}
after an expansion to second order in the gravitational fluctuation $g_{\mu\nu}\rightarrow g_{\mu\nu}+h_{\mu\nu}$ becomes \cite{ArkaniHamed:2008yf}
\begin{equation}\label{bulkkk}
S_{\rm bulk}=-\frac{1}{4}\frac{1}{2\kappa^2}\int d^{5}x\sqrt{-g}(g^{\mu\nu}g^{\lambda\rho}g^{\sigma\tau}\partial_{\mu}h_{\lambda\sigma}(x,u)\partial_{\nu}h_{\rho\tau}(x,u)-\frac{1}{2}g^{\mu\nu}\partial_{\mu}h(x,u)\partial_{\nu}h(x,u)-2R^{\lambda\rho\sigma\tau}h_{\lambda\sigma}h_{\rho\tau}),
\end{equation}
where the gravitational coupling $\frac{1}{2\kappa^{2}}=\frac{1}{16\pi G}$, $h\equiv g^{\mu\nu}h_{\mu\nu}$ and $R^{\lambda\rho\sigma\tau}$ is the Riemann curvature tensor constructed out of the background metric $g_{\mu\nu}$ which has horizon at $u=1$ and boundary at $u=0$.

And, by choosing a gauge at which $h=0$, considering only the transversal component $h_{xy}(t,z,u)$, and keeping only the kinetic terms, the action (\ref{bulkkk}) simplifies to
 \begin{equation}\label{bulk1}
S_{\rm bulk1}=\int d^{5}x\mathcal{L}_{1}=-\frac{1}{2}\frac{1}{2\kappa^2}\int d^{5}x\sqrt{-g}g^{\mu\mu}g^{xx}g^{yy}\partial_{\mu}h_{xy}(t,z,u)\partial_{\mu}h_{xy}(t,z,u),
\end{equation}
which can also be written as
 \begin{equation}\label{bulk2}
S_{\rm bulk2}=\int d^{5}x\mathcal{L}_{2}=-\frac{1}{2}\frac{1}{2\kappa^2}\int d^{5}x\sqrt{-g}g^{\mu\mu}g_{xx}g_{yy}\partial_{\mu}h^{xy}(t,z,u)\partial_{\mu}h^{xy}(t,z,u),
\end{equation}
and
\begin{equation}\label{bulk3}
S_{\rm bulk3}=\int d^{5}x\mathcal{L}_{3}=-\frac{1}{2}\frac{1}{2\kappa^2}\int d^{5}x\sqrt{-g}g^{\mu\mu}g^{xx}g_{yy}\partial_{\mu}h_x^{y}(t,z,u)\partial_{\mu}h_x^{y}(t,z,u).
\end{equation}
It's easy to see from (\ref{bulk3}) that for isotropic spacetime, where $g^{xx}g_{yy}=1$, $S_{\rm bulk3}$ is similar to an effective action for a scalar field $\phi_{3}=h_x^{y}$ with constant gravitational coupling $\frac{1}{2\kappa^2}$ while from (\ref{bulk1}) and (\ref{bulk2}) we see that (\ref{bulk1}) and (\ref{bulk2}) are similar to effective actions for scalar fields $\phi_{1}=h_{xy}$ and $\phi_{2}=h^{xy}$ with $u$ dependent effective couplings $\frac{1}{2\kappa^2}g^{xx}(u)g^{yy}(u)$ and $\frac{1}{2\kappa^2}g_{xx}(u)g_{yy}(u)$, respectively.

Varying the above bulk actions (\ref{bulk1}), (\ref{bulk2}), and (\ref{bulk3}) give us the equations of motion for the gravitational fluctuations $h_{xy}(t,z,u)$, $h^{xy}(t,z,u)$ and $h_x^{y}(t,z,u)$, respectively, as
\begin{equation}\label{isoeom1}
 \partial_{\mu}(\sqrt{-g}g^{\mu\mu}g^{xx}g^{yy}\partial_{\mu}h_{xy}(t,z,u))=0,
 \end{equation}
\begin{equation}\label{isoeom2}
 \partial_{\mu}(\sqrt{-g}g^{\mu\mu}g_{xx}g_{yy}\partial_{\mu}h^{xy}(t,z,u))=0,
 \end{equation}
and
\begin{equation}\label{isoeom3}
 \partial_{\mu}(\sqrt{-g}g^{\mu\mu}g^{xx}g_{yy}\partial_{\mu}h_x^y(t,z,u))=0.
 \end{equation}

\section{\label{paradigm}Membrane paradigm}
Integrating by parts the bulk actions (\ref{bulk1}), (\ref{bulk2}), and (\ref{bulk3}), and using the corresponding equations of motion (\ref{isoeom1}), (\ref{isoeom2}), and (\ref{isoeom3}), we'll be left with the on-shell boundary actions
\begin{equation}\label{ons1}
S_{on-shell1}=-S_{B1}[\epsilon],
\end{equation}
\begin{equation}\label{ons2}
S_{on-shell2}=-S_{B2}[\epsilon],
\end{equation}
and
\begin{equation}\label{ons3}
S_{on-shell3}=-S_{B3}[\epsilon],
\end{equation}
where the boundary actions at $u=\epsilon$, $S_{B1}[\epsilon]$, $S_{B2}[\epsilon]$, and $S_{B3}[\epsilon]$, are given by
\begin{equation}\label{baction1}
S_{B1}[\epsilon]=-\frac{1}{2}\frac{1}{2\kappa^2}\int_{u=\epsilon}d^{4}x\sqrt{-g}g^{uu}g^{xx}g^{yy}h_{xy}(t,z,u)\partial_{u}h_{xy}(t,z,u),
\end{equation}
\begin{equation}\label{baction2}
S_{B2}[\epsilon]=-\frac{1}{2}\frac{1}{2\kappa^2}\int_{u=\epsilon}d^{4}x\sqrt{-g}g^{uu}g_{xx}g_{yy}h^{xy}(t,z,u)\partial_{u}h^{xy}(t,z,u),
\end{equation}
and
\begin{equation}\label{baction3}
S_{B3}[\epsilon]=-\frac{1}{2}\frac{1}{2\kappa^2}\int_{u=\epsilon}d^{4}x\sqrt{-g}g^{uu}g^{xx}g_{yy}h_x^y(t,z,u)\partial_{u}h_x^y(t,z,u).
\end{equation}
Then, by imposing the boundary conditions that the conjugate momenta in the bulk $\Pi^{uxy}=\frac{\partial\mathcal{L}_{1}}{\partial \partial_{u}h_{xy}}$ or $\Pi^{u}_{xy}=\frac{\partial\mathcal{L}_{2}}{\partial \partial_{u}h^{xy}}$ or $\Pi^{ux}\,_{y}=\frac{\partial\mathcal{L}_{3}}{\partial \partial_{u}h^{y}_{x}}$ should be equal to the momentum currents $T^{xy}=\frac{\delta S_{B1}}{\delta h_{xy}}$ or $T_{xy}=\frac{\delta S_{B2}}{\delta h^{xy}}$ or $T^{x}_{y}=\frac{\delta S_{B3}}{\delta h^{y}_{x}}$, respectively, at the boundary or at any other hypersurface at $u=\epsilon$, we can calculate the corresponding momentum currents as
 \begin{equation}\label{t1}
 T^{xy}=\frac{\delta S_{B1}}{\delta h_{xy}}=\frac{\partial\mathcal{L}_{1}}{\partial \partial_{u}h_{xy}}=-\frac{1}{2\kappa^2}\sqrt{-g}g^{uu}g^{xx}g^{yy}\partial_{u}h_{xy}(t,z,u),
 \end{equation}
 \begin{equation}\label{t2}
 T_{xy}=\frac{\delta S_{B2}}{\delta h^{xy}}=\frac{\partial\mathcal{L}_{2}}{\partial \partial_{u}h^{xy}}=-\frac{1}{2\kappa^2}\sqrt{-g}g^{uu}g_{xx}g_{yy}\partial_{u}h^{xy}(t,z,u),
 \end{equation}
 and
 \begin{equation}\label{t3}
 T^{x}\,_{y}=\frac{\delta S_{B3}}{\delta h^{y}_{x}}=\frac{\partial\mathcal{L}_{3}}{\partial \partial_{u}h^{y}_{x}}=-\frac{1}{2\kappa^2}\sqrt{-g}g^{uu}g^{xx}g_{yy}\partial_{u}h_x^y(t,z,u).
 \end{equation}

According to the membrane paradigm \cite{Kovtun:2003wp, Iqbal:2008by}, in order to evaluate the shear viscosities, it's enough to evaluate the conjugate momenta $\Pi^{uxy}$, $\Pi^{u}_{xy}$ and $\Pi^{ux}\,_{y}$ at the horizon $u=1$ where we can use the Eddington-Finklestein coordinate $v$ defined by
\begin{equation}
  dv=dt-\sqrt{\frac{g_{uu}}{-g_{tt}}}du=0,
\end{equation}
to re-write
\begin{equation}
\partial_{u}h_{xy}(t,z,u)=\sqrt{\frac{g_{uu}}{-g_{tt}}}\partial_{t}h_{xy}(t,z,u),
\end{equation}
\begin{equation}
\partial_{u}h^{xy}(t,z,u)=\sqrt{\frac{g_{uu}}{-g_{tt}}}\partial_{t}h^{xy}(t,z,u),
\end{equation}
and
\begin{equation}
\partial_{u}h_x^y(t,z,u)=\sqrt{\frac{g_{uu}}{-g_{tt}}}\partial_{t}h_x^y(t,z,u).
\end{equation}
Therefore, the momentum currents or conjugate momenta (\ref{t1}), (\ref{t2}), and (\ref{t3}) at the horizon $u=1$ become
 \begin{equation}\label{t11}
 T^{xy}=-\frac{1}{2\kappa^2}\sqrt{g_{xx}(1)g_{yy}(1)g_{zz}(1)}g^{xx}(1)g^{yy}(1)\partial_{t}h_{xy}(t,z,u=1),
 \end{equation}
 \begin{equation}\label{t22}
 T_{xy}=-\frac{1}{2\kappa^2}\sqrt{g_{xx}(1)g_{yy}(1)g_{zz}(1)}g_{xx}(1)g_{yy}(1)\partial_{t}h^{xy}(t,z,u=1),
 \end{equation}
 and
 \begin{equation}\label{t33}
 T^{x}\,_{y}=-\frac{1}{2\kappa^2}\sqrt{g_{xx}(1)g_{yy}(1)g_{zz}(1)}g^{xx}(1)g_{yy}(1)\partial_{t}h_x^y(t,z,u=1).
 \end{equation}
Since, we can write the momentum currents (or the response for a change in $h_{xy}(t,z,u=1)$ or $h^{xy}(t,z,u=1)$ or $h_x^y(t,z,u=1)$) in terms of the shear viscosities $\eta^{xyxy}$, $\eta_{xyxy}$, and $ \eta^{x}\,_{y}\,^{x}\,_{y}$ as (see for example \cite{Baier:2007ix})
 \begin{equation}\label{t111}
 T^{xy}=-\eta^{xyxy}\partial_{t}h_{xy}(t,z,u=1),
 \end{equation}
  \begin{equation}\label{t222}
 T_{xy}=-\eta_{xyxy}\partial_{t}h^{xy}(t,z,u=1),
 \end{equation}
 and
 \begin{equation}\label{t333}
 T^{x}\,_{y}=-\eta^{x}\,_{y}\,^{x}\,_{y}\partial_{t}h_x^y(t,z,u=1),
 \end{equation}
we can compare (\ref{t111}), (\ref{t222}), and (\ref{t333}) with (\ref{t11}), (\ref{t22}), and (\ref{t33}), respectively, to infer that
\begin{equation}
  \eta^{xyxy}=\frac{1}{2\kappa^2}\sqrt{g_{xx}(1)g_{yy}(1)g_{zz}(1)}g^{xx}(1)g^{yy}(1),
\end{equation}
\begin{equation}
  \eta_{xyxy}=\frac{1}{2\kappa^2}\sqrt{g_{xx}(1)g_{yy}(1)g_{zz}(1)}g_{xx}(1)g_{yy}(1),
\end{equation}
 and
 \begin{equation}
  \eta^{x}\,_{y}\,^{x}\,_{y}=\frac{1}{2\kappa^2}\sqrt{g_{xx}(1)g_{yy}(1)g_{zz}(1)}g^{xx}(1)g_{yy}(1).
 \end{equation}

Finally, since the entropy density $s=\frac{S}{V}$ is given by
\begin{equation}
  s=\frac{1}{4G}\frac{A}{V}=\frac{1}{4G}\sqrt{g_{xx}(1)g_{yy}(1)g_{zz}(1)},
\end{equation}
the shear viscosity to entropy densities $\frac{\eta^{xyxy}}{s}$, $\frac{\eta_{xyxy}}{s}$ and $\frac{\eta^{x}\,_{y}\,^{x}\,_{y}}{s}$ become
\begin{equation}\label{eta11}
\frac{\eta^{xyxy}}{s}=\frac{4G}{2\kappa^2}g^{xx}(1)g^{yy}(1)=\frac{1}{4\pi}g^{xx}(1)g^{yy}(1),
\end{equation}
\begin{equation}\label{eta22}
\frac{\eta_{xyxy}}{s}=\frac{4G}{2\kappa^2}g_{xx}(1)g_{yy}(1)=\frac{1}{4\pi}g_{xx}(1)g_{yy}(1),
\end{equation}
and
\begin{equation}\label{eta33}
\frac{\eta^{x}\,_{y}\,^{x}\,_{y}}{s}=\frac{4G}{2\kappa^2}g^{xx}(1)g_{yy}(1)=\frac{1}{4\pi}g^{xx}(1)g_{yy}(1),
\end{equation}
where we used $2\kappa^2=16\pi G$ to get the last lines. Equations (\ref{eta11}), (\ref{eta22}), and (\ref{eta33}) are the main results of this paper. And, one can immediately see that for isotropic spacetime where $g^{xx}(1)g_{yy}(1)=1$, $\frac{\eta^{x}\,_{y}\,^{x}\,_{y}}{s}$ takes the universal value $\frac{1}{4\pi}$ but $\frac{\eta^{xyxy}}{s}$ and $\frac{\eta_{xyxy}}{s}$ are non-universal and depend on the details of the bulk metric $g_{ij}$.

\section{\label{applications}Applications}
We'll use the equations (\ref{eta11}), (\ref{eta22}), and (\ref{eta33}) in order to calculate the shear viscosity to entropy density ratios $\frac{\eta^{xyxy}}{s}$, $\frac{\eta_{xyxy}}{s}$, and $\frac{\eta^{x}\,_{y}\,^{x}\,_{y}}{s}$ of the strongly coupled $\mathcal{N}=4$ super-Yang-Mills (SYM), ABJM, and (2,0) superconformal plasmas.
\subsection{D3-branes}
The supergravity dual to strongly coupled $\mathcal{N}=4$ SYM living on the 4-dimensional world volume of $N_{c}$ D3-branes is studied in asymptotically $AdS_5$ bulk spacetime \cite{Kovtun:2003wp}
\begin{eqnarray}\label{ads5}
ds^{2}&=&g_{\mu\nu}dx^{\mu}dx^{\nu}={\pi^2 T^2 R^2\over u}\left( -f(u) dt^2 + dx^2 + dy^2 +dz^2\right)+{R^2\over 4 f(u) u^2} du^2 \ ,
\end{eqnarray}
where $T = \frac{r_0}{\pi R^2}$ is the Hawking temperature, $R^{4}=\lambda\ell_{s}^{4}$, $\lambda=g^{2}_{YM}N_{c}$ is the 't Hooft coupling, we've introduced $u = r_0^2/r^2$, $f(u)=1-u^2$, the horizon corresponds to $u=1$, and the boundary to $u=0$.

Using (\ref{ads5}) in (\ref{eta11}), (\ref{eta22}), and (\ref{eta33}), we find that
\begin{equation}\label{11}
\frac{\eta^{xyxy}}{s}=\frac{1}{4\pi}g^{xx}(1)g^{yy}(1)=\frac{1}{4\pi}\frac{1}{(\pi TR)^4}\geq \frac{1}{4\pi},
\end{equation}
\begin{equation}\label{12}
\frac{\eta_{xyxy}}{s}=\frac{1}{4\pi}g_{xx}(1)g_{yy}(1)=\frac{1}{4\pi}(\pi TR)^4\leq\frac{1}{4\pi},
\end{equation}
and
\begin{equation}\label{n4}
\frac{\eta^{x}\,_{y}\,^{x}\,_{y}}{s}=\frac{1}{4\pi}g^{xx}(1)g_{yy}(1)=\frac{1}{4\pi},
\end{equation}
where we used the fact that $\pi TR\leq1$ in the extremal limit $r_{0}\leq R$. We see that $\frac{\eta^{x}\,_{y}\,^{x}\,_{y}}{s}$ takes the universal value $\frac{1}{4\pi}$ as expected but $\frac{\eta^{xyxy}}{s}$ and $\frac{\eta_{xyxy}}{s}$ are non-universal and depend on both the temperature $T$ and some length scale $R$ which doesn't have any physical meaning in $\mathcal{N}=4$ SYM plasma on flat space. So, for $\mathcal{N}=4$ SYM plasma on flat space, the correct choice for the shear viscosity to entropy density ratio must be the one independent of $R$ which is the universal one $\frac{\eta^{x}\,_{y}\,^{x}\,_{y}}{s}$ (\ref{n4}) but in other boundary theories where $R$ has physical meaning, like $\mathcal{N}=4$ SYM plasma on a sphere of radius $R$, the non-universal ones $\frac{\eta^{xyxy}}{s}$ and $\frac{\eta_{xyxy}}{s}$ might play a bigger role and be identified as the correct shear viscosities.

\subsection{M2-branes}
The gravity dual to strongly coupled ABJM plasma living on the 3-dimensional world volume of $N$ M2-branes is studied in asymptotically $AdS_4$ bulk spacetime \cite{Kovtun:2003wp}
\begin{eqnarray}\label{ads4}
ds^{2}&=&g_{\mu\nu}dx^{\mu}dx^{\nu}={(\frac{2}{3})^2\pi^2 T^2 R^2\over u^4}\left( -f(u) dt^2 + dx^2 + dy^2\right)+{R^2\over f(u) u^2} du^2 \ ,
\end{eqnarray}
where $T = \frac{r_0^2}{\frac{2}{3}\pi R^3}$ is the Hawking temperature, $R^{6}=\frac{1}{2}\pi^2N\ell_{p}^{6}$ is the radius of the $AdS_{4}$ space, $\ell_{p}$ is Planck's length, $u = r_0/r$, $f(u)=1-u^6$, the horizon corresponds to $u=1$, and the boundary to $u=0$.

So, using (\ref{ads4}) in (\ref{eta11}), (\ref{eta22}), and (\ref{eta33}), we find that
\begin{equation}\label{21}
\frac{\eta^{xyxy}}{s}=\frac{1}{4\pi}g^{xx}(1)g^{yy}(1)=\frac{1}{4\pi}\frac{1}{(\frac{2}{3}\pi TR)^4}\geq \frac{1}{4\pi},
\end{equation}
\begin{equation}\label{22}
\frac{\eta_{xyxy}}{s}=\frac{1}{4\pi}g_{xx}(1)g_{yy}(1)=\frac{1}{4\pi}(\frac{2}{3}\pi TR)^4\leq \frac{1}{4\pi},
\end{equation}
and
\begin{equation}
\frac{\eta^{x}\,_{y}\,^{x}\,_{y}}{s}=\frac{1}{4\pi}g^{xx}(1)g_{yy}(1)=\frac{1}{4\pi},
\end{equation}
where we used the fact that $\frac{2}{3}\pi TR\leq1$ in the extremal limit $r_{0}\leq R$. And, similar to the $\mathcal{N}=4$ SYM plasma on flat space, the physical choice for the shear viscosity of the ABJM plasma on flat space should be $\frac{\eta^{x}\,_{y}\,^{x}\,_{y}}{s}$.

\subsection{M5-branes}
The gravity dual to strongly coupled (2,0) superconformal plasma living on the 6-dimensional world volume of $N$ M5-branes is studied in asymptotically $AdS_7$ bulk spacetime \cite{Kovtun:2003wp}
\begin{equation}\label{ads7}
ds^{2}=g_{\mu\nu}dx^{\mu}dx^{\nu}={\frac{16}{9}\pi^2 T^2 R^2\over u}\left( -f(u) dt^2 + dx^2 + dy^2 + dz^2 + dw^2 + dv^2\right)+{R^2\over f(u) u^2} du^2 \ ,
\end{equation}
where $T = \frac{r_0^{\frac{1}{2}}}{\frac{4}{3}\pi R^{\frac{3}{2}}}$ is the Hawking temperature, $R^{3}=8\pi^2N\ell_{p}^{3}$ is the radius of the $AdS_{7}$ space, $\ell_{p}$ is Planck's length, $u = r_0/r$, $f(u)=1-u^6$, the horizon corresponds to $u=1$, and the boundary to $u=0$.

So, using (\ref{ads7}) in (\ref{eta11}), (\ref{eta22}), and (\ref{eta33}), we find that
\begin{equation}\label{31}
\frac{\eta^{xyxy}}{s}=\frac{1}{4\pi}g^{xx}(1)g^{yy}(1)=\frac{1}{4\pi}\frac{1}{(\frac{4}{3}\pi TR)^4}\geq \frac{1}{4\pi},
\end{equation}
\begin{equation}\label{32}
\frac{\eta_{xyxy}}{s}=\frac{1}{4\pi}g_{xx}(1)g_{yy}(1)=\frac{1}{4\pi}(\frac{4}{3}\pi TR)^4\leq \frac{1}{4\pi},
\end{equation}
and
\begin{equation}
\frac{\eta^{x}\,_{y}\,^{x}\,_{y}}{s}=\frac{1}{4\pi}g^{xx}(1)g_{yy}(1)=\frac{1}{4\pi},
\end{equation}
where we used the fact that $\frac{4}{3}\pi TR\leq1$ in the extremal limit $r_{0}\leq R$. And, similar to the $\mathcal{N}=4$ SYM and ABJM plasmas on flat space, the physical choice for the shear viscosity of the (2,0) superconformal theory on flat space should be $\frac{\eta^{x}\,_{y}\,^{x}\,_{y}}{s}$.

\section{Conclusion}
We've revisited the holographic calculations of the shear viscosity tensor to entropy density ratios of strongly coupled isotropic plasmas and have found that $\frac{\eta^{x}\,_{y}\,^{x}\,_{y}}{s}$ (\ref{eta33}) takes the universal value $\frac{1}{4\pi}$ while $\frac{\eta^{xyxy}}{s}$ (\ref{eta11}) and $\frac{\eta_{xyxy}}{s}$ (\ref{eta22}) are non-universal and depend on the details of the background metric $g_{ij}$.

We've also applied the formulae for $\frac{\eta^{xyxy}}{s}$ (\ref{eta11}) and $\frac{\eta_{xyxy}}{s}$ (\ref{eta22}) to strongly coupled $\mathcal{N}=4$ SYM (\ref{11})(\ref{12}), ABJM (\ref{21})(\ref{22}), and (2,0) superconformal (\ref{31})(\ref{32}) plasmas on flat spaces and have found that they depend explicitly on the temperature $T$ and radius $R$ of the AdS spacetime.

The AdS radius $R$ has no physical meaning for the superconformal gauge theories on flat spaces hence $\frac{\eta^{xyxy}}{s}$ and $\frac{\eta_{xyxy}}{s}$ can be discarded as unphysical. However, when the gauge theories live on compact spaces like spheres with radius $R$ (which is also the radius $R$ of the AdS spacetime), $\frac{\eta^{xyxy}}{s}$ and $\frac{\eta_{xyxy}}{s}$ can be considered physical and their physical implications for the strongly coupled isotropic plasmas on compact spaces (like spheres with radius R) should be investigated further in the future, as we'll in \cite{Mamo:2013}.

%\acknowledgments
%The author thanks A. Lewis Licht, Bo Ling, and Misha Stephanov for reading the draft.

\bibliographystyle{JHEP}
\bibliography{ref}

\end{document}